\documentclass[12pt]{article}
\usepackage{epsfig}

\def\beq{\begin{equation}}
\def\eeq#1{\label{#1}\end{equation}}
\def\eeqn{\end{equation}}

\def\beqa{\begin{eqnarray}}
\def\eeqa#1{\label{#1}\end{eqnarray}}
\def\eeqan{\end{eqnarray}}

\let\bar=\overbar

\def\Dslash{\not{\hbox{\kern-4pt $D$}}}
\def\dslash{\not{\hbox{\kern-2pt $\del$}}}

\def\msb{{\bar{\ssstyle M \kern -1pt S}}}

\def\Title#1{\begin{center} {\Large {\bf #1} } \end{center}}

\begin{document}

\Title{Short-baseline Neutrinos:\\Recent Results and Future Prospects}

\bigskip\bigskip


\begin{raggedright}  

{\it Ryan B. Patterson\index{Patterson, R. B.}\\
Physics Department\\
California Institute of Technology\\
Pasadena, California 91125\\
U.S.A.}
\bigskip\bigskip
\end{raggedright}

\section{Oscillations}\index{neutrino oscillations}
The evidence is compelling that neutrinos undergo flavor change as they propagate.  In recent years, experiments have observed the phenomenon of neutrino oscillations using disparate neutrino sources:\ the sun, fission reactors, accelerators, and secondary cosmic rays.  The standard model of particle physics needs only simple extensions -- neutrino masses and mixing -- to accommodate all neutrino oscillation results to date, save one.  The 3.8$\sigma$-significant $\bar{\nu}_e$ excess reported by the LSND \index{LSND} collaboration~\cite{LSND} is consistent with $\bar{\nu}_\mu\,\mathord{\rightarrow}\,\bar{\nu}_e$ oscillations with a mass-squared splitting of $\Delta m^2\,\mathord{\sim}\, 1\ \mathrm{eV}^2$.  This signal, which has not been independently verified, is inconsistent with other oscillation evidence unless more invasive standard model extensions ({\em e.g.}, sterile neutrinos) are considered.

The Mini Booster Neutrino Experiment (MiniBooNE) \index{MiniBooNE} is designed to search for $\nu_\mu\,\mathord{\rightarrow}\,\nu_e$ oscillations with sufficient sensitivity to confirm or refute the LSND signal.  MiniBooNE uses the Fermilab Booster neutrino beam, which begins with 8~GeV protons impinging on a beryllium target.  The positively charged secondary mesons (mostly $\pi^+$, but some $K^+$) produced in the target are magnetically focused forward into a 50~m air-filled decay region.  Their subsequent decay chains lead to the high intensity $\mathord{\sim}1~\mathrm{GeV}$ neutrino source.  The neutrinos are predominantly $\nu_\mu$, but $K$ and $\mu$ decays lead to a $0.6\%$ $\nu_e$ contamination that represents a large irreducible background to the $\nu_\mu\,\mathord{\rightarrow}\,\nu_e$ search (as the best-fit LSND oscillation probability is $\sim$$0.3$\%).

The MiniBooNE detector sits 541~m downstream of the proton target, with most of the space in between occupied by earth.  The detector is a 6.1~m radius spherical steel tank filled with 800 tons of mineral oil.  An opaque shell of diameter 5.75~m, concentric with the steel tank, divides the oil into two optically isolated regions.  The thin outer region is instrumented with 240 8-inch PMTs and serves as a veto shield for incoming cosmic rays and for partially contained neutrino events.  The inner main region is viewed by 1280 8-inch PMTs.  The Cherenkov (and, at a lower level, scintillation) photons produced by high energy charged particles in the mineral oil lead to light patterns on the PMT array.

Charged current neutrino interactions in the MiniBooNE detector are identified by the characteristic ring of Cherenkov light created by the outgoing charged lepton.  A muon's long, straight track leads to a sharp ring that fills in as the muon ranges out.  An electron induces an electromagnetic shower, leading to a diffuse, but still ring-like, pattern.

High energy photons, such as those coming from $\pi^0$ decay, also produce electron-like ring patterns.  Neutral current production of $\pi^0$'s, usually proceeding through a $\Delta$ resonance, leads to the largest misidentification background in MiniBooNE.  This misidentification occurs when one of the two photons from the $\pi^0$ decay goes unnoticed, either by having too little energy or by having its ring pattern obscured by the other's.  In such cases, the event appears to contain a single electromagnetic shower and will be classified as a $\nu_e$ charged current event.  A related photon-based background comes from the rare ($\mathord{\sim}0.5\%$) electromagnetic decay $\Delta\rightarrow N\gamma$.

\begin{figure}[htb]
\begin{center}
\epsfig{file=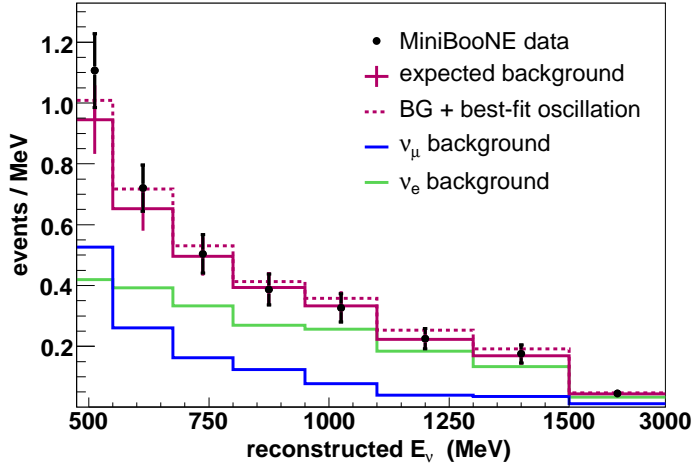,width=0.65\textwidth}
\caption{Reconstructed neutrino energy spectrum for the MiniBooNE $\nu_e$ candidate sample.  The black points show the data with statistical error bars.  The magenta curves show the expected background-only spectrum (solid) and best-fit oscillation spectrum (dashed).  The separate $\nu_\mu$ and $\nu_e$ components of the background are also shown (blue and green).  The null and best-fit scenarios are insignificantly different ($\Delta\chi^2=0.83$).}
\label{fig:mbspec}
\end{center}
\end{figure}

\begin{figure}[htb]
\begin{center}
\epsfig{file=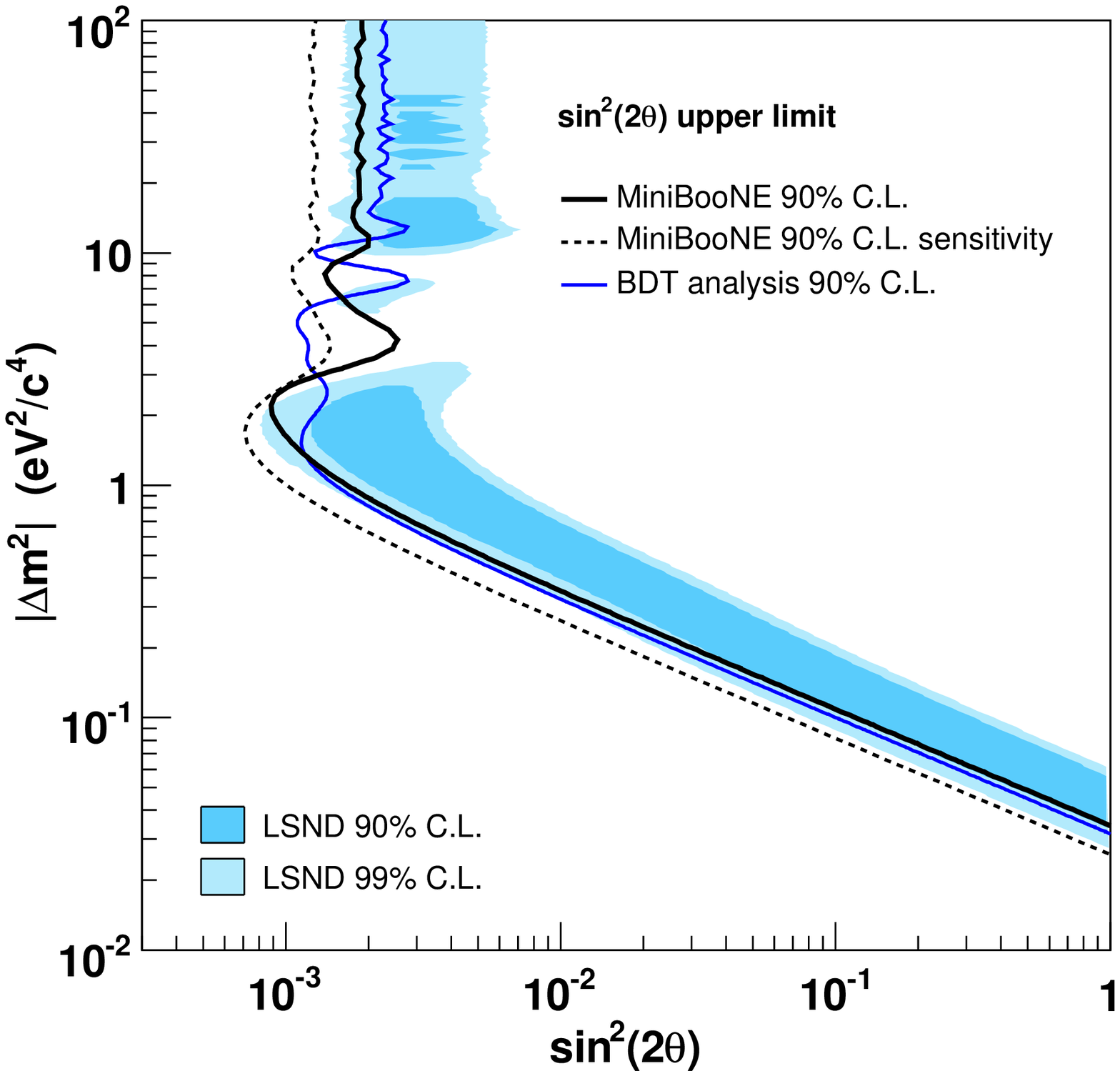,width=0.65\textwidth}
\caption{MiniBooNE exclusion contours. The filled regions show the LSND 90\% and 99\% C.L.\ allowed regions in oscillation parameter space $(\sin^2(2\theta),\Delta m^2)$.  MiniBooNE excludes at 90\% C.L.\ the oscillation parameters above the thick black line.  Also shown are the MiniBooNE sensitivity contour (dashed) and the limit contour from a second, fairly independent analysis (thin blue).}
\label{fig:mbcontour}
\end{center}
\end{figure}

Further analysis details can be found in the first MiniBooNE oscillation paper, published in early 2007~\cite{mbosc}.  Using a data sample of $1.7\mathord{\times}10^{6}$ neutrino interactions from $5.58\mathord{\times}10^{20}$~protons-on-target, MiniBooNE found  no evidence for LSND-like $\nu_\mu\,\mathord{\rightarrow}\,\nu_e$ oscillations, neither in the count of $\nu_e$ candidates [data: 380, expectation: $358\pm 19_{\mathrm{stat}}\pm 35_{\mathrm{syst}}$] nor in the shape of the neutrino energy spectrum.  The spectrum is shown in Figure~\ref{fig:mbspec}, and the MiniBooNE limit contour for two-neutrino mixing parameters is shown in Figure~\ref{fig:mbcontour}.  MiniBooNE also published a combined analysis of Bugey, KARMEN2, LSND, and MiniBooNE data, concluding that the four experiments are mutually compatible at only 3.9\% C.L.~\cite{mbcombined}

The MiniBooNE oscillation analysis used a low-energy threshold of 475~MeV.  A look below this threshold, down to 300~MeV, reveals a $3.7\sigma$ discrepancy between data and expectation.  The data excess is not consistent with two-neutrino oscillations.  About 25\% of the excess has been explained by the lack of photonuclear absorption in the simulation.  (This process can remove one of the $\pi^0\rightarrow\gamma\gamma$ photons from an event, leaving behind an electron-like signature.)  The remainder of the excess is as yet unexplained.  Anomaly-mediated $\gamma$ production was noted in Ref.~\cite{hhh} as a potential solution, but the rate and photon spectrum for this standard model process have not yet been well determined.

Although MiniBooNE was the higher profile experiment, the CHORUS \index{CHORUS} collaboration also presented short-baseline oscillation results this year.  CHORUS searched for $\nu_\mu\,\mathord{\rightarrow}\,\nu_\tau$ oscillations at high $\Delta m^2$ ($\mathord{>}1~\mathrm{eV}^2$) by looking for $\tau$ lepton appearance in a 770~kg emulsion detector exposed to a 26~GeV broadband $\nu_\mu$ beam.  The vertex detection provided by the emulsion was supplemented by a fiber tracker, hadron and muon spectrometers, and hadronic and electromagnetic calorimeters.

\begin{figure}[htb]
\begin{center}
\epsfig{file=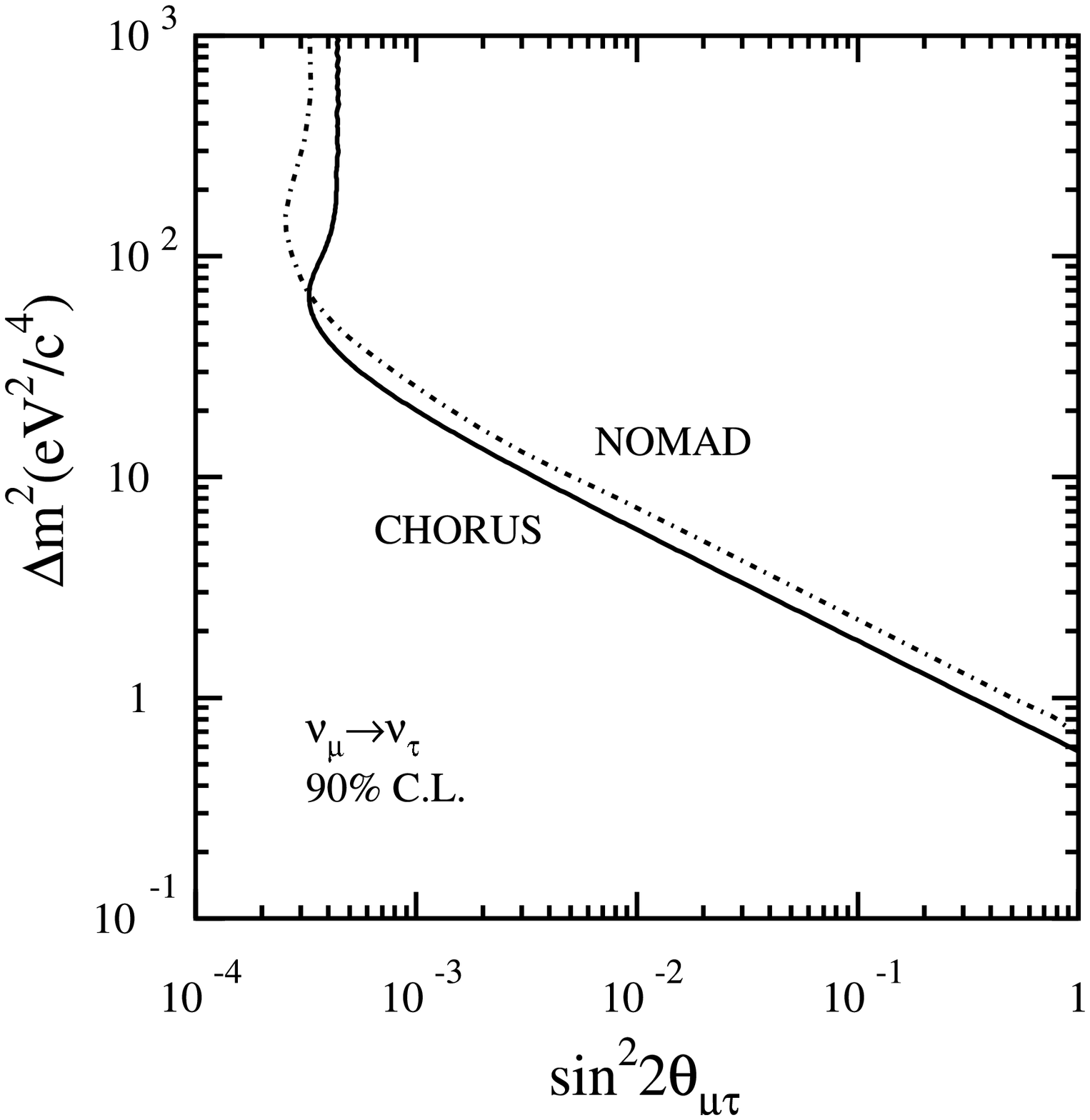,width=0.6\textwidth}
\caption{90\% C.L.\ limit in oscillation parameter space established by CHORUS.  Points above the curve are excluded.  The similar NOMAD limit is also shown.  Figure from Ref.~\cite{chorus}.}
\label{fig:chorus}
\end{center}
\end{figure}

The recent paper reports an updated analysis of the complete 1994-1997 CHORUS data set using improved event reconstruction algorithms and a new automated system for scanning the emulsion plates~\cite{chorus}.  No $\nu_\tau$ appearance was observed, and the oscillation limits were improved by 30\% over their earlier result.  Figure~\ref{fig:chorus} shows the oscillation parameters excluded by CHORUS.

\section{Cross sections}
\begin{figure}[htb]
\begin{center}
\epsfig{file=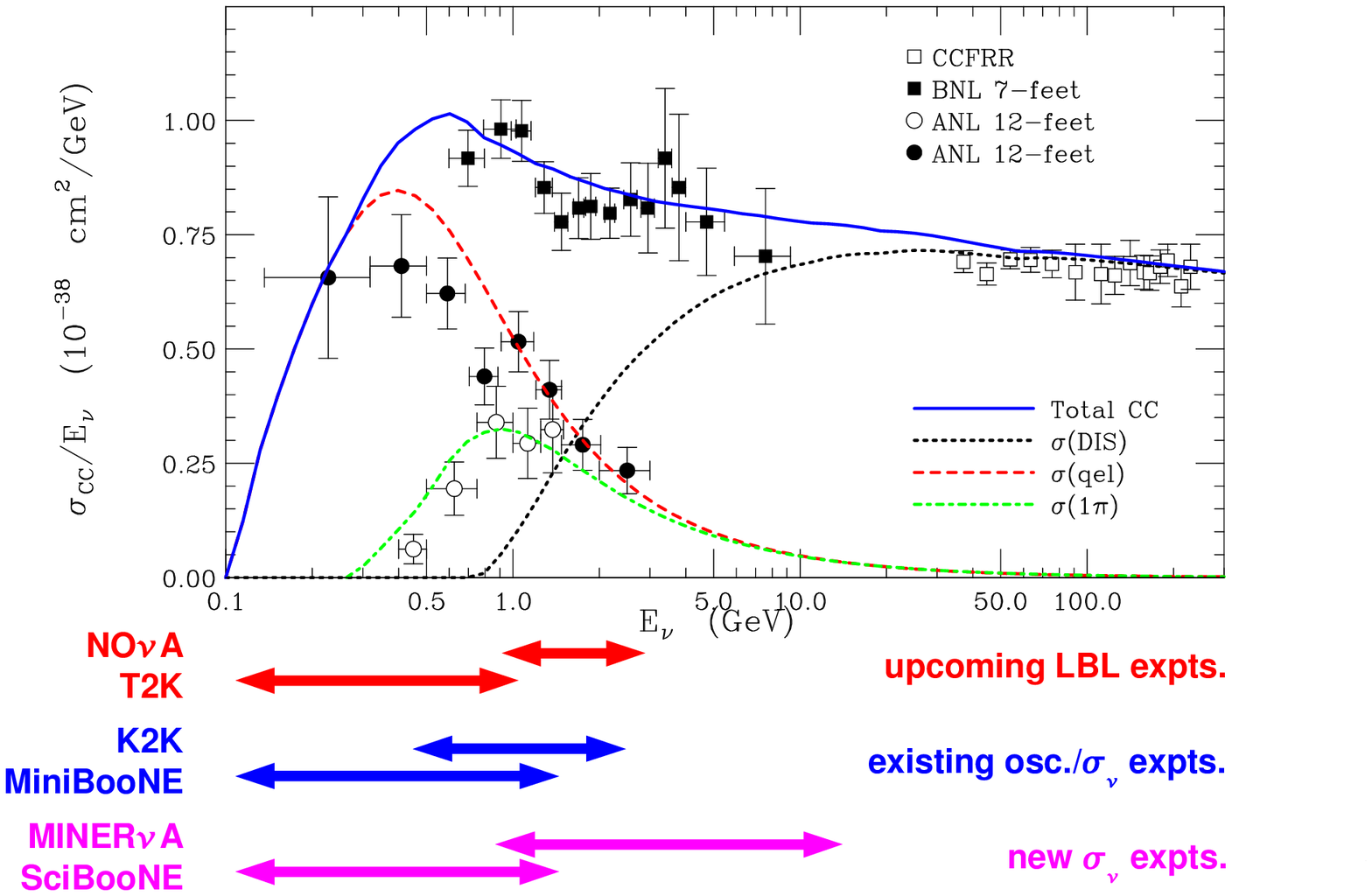,width=0.8\textwidth}
\caption{Charged current $\nu$ cross section data.  The top portion of the figure is taken from Ref.~\cite{lipari}.  The upcoming long-baseline experiments NO$\nu$A and T2K will benefit from improved low energy cross section measurements.  As discussed in the text, K2K and MiniBooNE, both oscillation experiments, are providing additional data, but the next big steps in cross section precision will come with MINER$\nu$A and SciBooNE.}
\label{fig:ediag}
\end{center}
\end{figure}

\index{neutrino cross sections}
The continuing surge of neutrino oscillation experiments brings an urgent need for precision neutrino cross sections.  Cross sections for many relevant channels and energies have never been measured, and those that have are often decades old and imprecise.  Further, existing measurements are rarely on nuclear targets.  To give a sense of the situation, Figure~\ref{fig:ediag} shows high-statistics cross section data for charged current $\nu$ scattering over a wide energy range.  Lest the picture look too rosy, neutral current data below 3~GeV are nearly non-existent, with only a handful of usually spectrum-specific measurements available.

The K2K collaboration \index{K2K} is providing some help.  K2K is a long-baseline oscillation experiment that uses a 1.3~GeV broadband $\nu_\mu$ beam (97\% purity) created at KEK in Tsukuba, Japan, and directed toward the Super-Kamiokande detector 250~km away.  A near detector on the KEK site measures the neutrino flux for the oscillation search and, relevant for this note, provides a large sample of neutrino interactions from which cross sections can be extracted.  The near detector has four components: a 1~kton water Cherenkov detector, a scintillating fiber tracker (``SciFi'', water target), a scintillator bar tracker (``SciBar'', CH target), and a muon range stack (Fe).  The scintillator tracker has been replaced by a lead glass array, but no cross section results have yet come from the latter.

A generic feature of conventional GeV-scale neutrino beams is that the absolute rate of neutrinos, and often the energy spectrum, is poorly known.  This stems from the lack of quality cross section data for the production of pions in the beam targets.  K2K therefore reports ratios of neutrino cross sections, with the reference channel being either inclusive or quasi-elastic charged current scattering.

\begin{figure}[htb]
\begin{center}
\epsfig{file=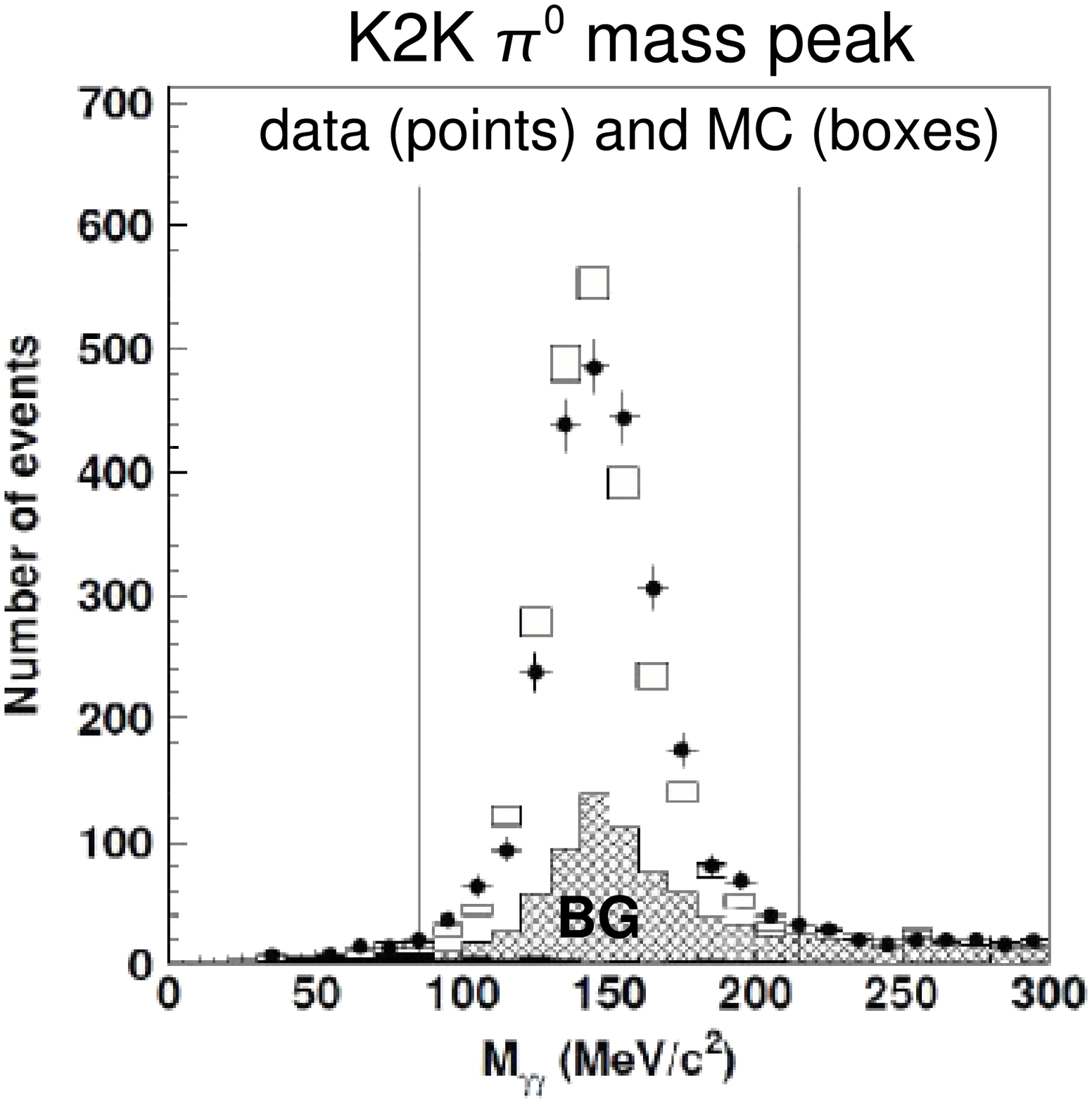,width=0.6\textwidth}
\caption{Data and Monte Carlo expectation for the $\pi^0$ mass peak reconstructed from neutral current single-$\pi^0$ events in the K2K 1-kton water Cherenkov detector.  The integrals between the vertical lines agree well within errors.  Figure adapted from Ref.~\cite{k2kncpi0}.}
\label{fig:k2kncpi0}
\end{center}
\end{figure}

\begin{figure}[htb]
\begin{center}
\epsfig{file=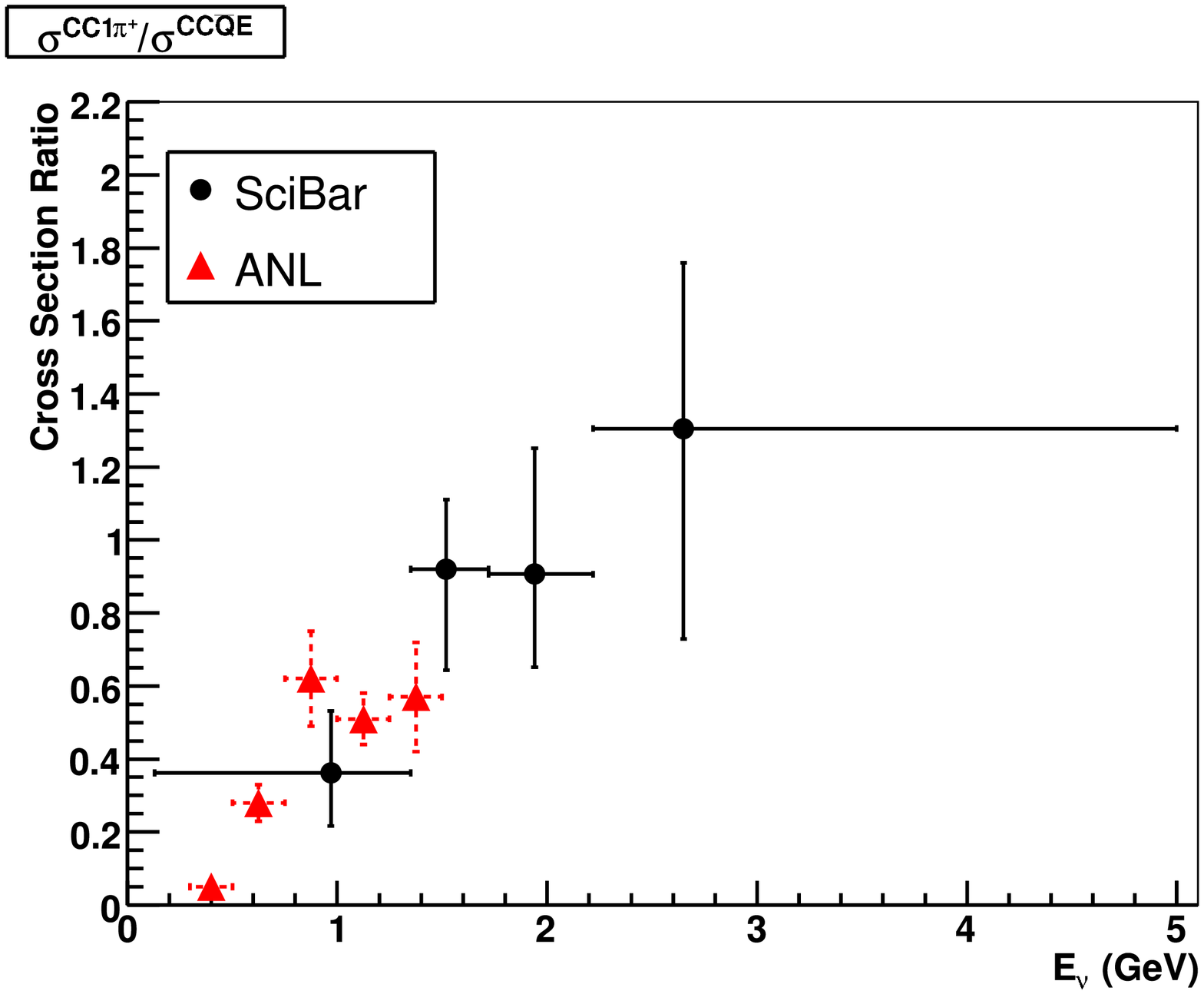,width=0.6\textwidth}
\caption{The ratio $\sigma_{\mathrm{CC1\pi^+}} / \sigma_{\mathrm{CCQE}}$ versus neutrino energy.  The black circles show the K2K SciBar measurement.  The red triangles show Argonne data taken from~\cite{anl1} and \cite{anl2}.  Figure taken from Ref.~\cite{k2kccpip}, which also includes a graph of data and Monte Carlo together showing excellent agreement.}
\label{fig:k2kccpip}
\end{center}
\end{figure}

Several recent K2K results relate to pion production\index{pion production}, and taken together they reveal a murky situation.  Figure~\ref{fig:k2kncpi0} shows the reconstructed $\pi^0$ mass peak from neutral current interactions in the near water Cherenkov detector.  From the event rate in the peak they extract the neutral current single-$\pi^0$ production cross section in ratio with the total charged current cross section~\cite{k2kncpi0}:
\begin{equation}
\sigma_{\mathrm{NC1\pi^0}} / \sigma_{\mathrm{CC}}=0.063\pm0.001_{\mathrm{stat}}\pm0.006_{\mathrm{stat}}\ .
\end{equation} 
This value agrees well with the Monte Carlo expectation of 0.064.  However, the analysis of {\em charged} current $\pi^0$ production in SciBar yields a number that is 40\% higher than the Monte Carlo prediction~\cite{k2kncpi0},
\begin{equation}
\sigma_{\mathrm{CC1\pi^0}} / \sigma_{\mathrm{CCQE}}=0.306\pm0.023_{\mathrm{stat}}\pm0.025_{\mathrm{stat}}\ .
\end{equation} 
Three things changed from the first measurement to the second: (1) neutral to charged current, (2) water to hydrocarbon target, (3) inclusive to quasi-elastic charged current reference channel.  Adding one further change ($\pi^+$ production rather than $\pi^0$) brings data and Monte Carlo {\em back into agreement}, although the errors are somewhat large~\cite{k2kccpip}.  The K2K $\sigma_{\mathrm{CC1\pi^+}} / \sigma_{\mathrm{CCQE}}$ data also agrees with previous Argonne bubble chamber data~\cite{k2kccpip}, as Figure~\ref{fig:k2kccpip} shows.  The upcoming dedicated cross section experiments discussed below are needed to clarify this situation.

MiniBooNE\index{MiniBooNE} has added to the pion production data set by publishing the first ever measurement of coherent $\pi^0$ production below 2~GeV.  In coherent production, the target nucleus (in this case, ${}^{12}\mathrm{C}$) stays intact and in the ground state after the interaction.  The MiniBooNE detector cannot observe the nuclear state directly, so the analysis uses the angular distribution (relative to the neutrino direction) of the outgoing $\pi^0$ to extract a coherent fraction.  (The detector also cannot resolve few-MeV de-excitation photons, so ``coherent'' here includes possible ${}^{12}\mathrm{C}^{*}$ final states.)  MiniBooNE finds that $(19.5 \pm 1.1_{\mathrm{stat}} \pm 2.5_{\mathrm{syst}})\%$ of its $\pi^0$ production is coherent, significantly below the Rein/Sehgal-based~\cite{reinsehgal} Monte Carlo expectation of 30\%~\cite{mbcoh}.  Preliminary $\overline{\nu}$ data shows a similar coherent $\pi^0$ discrepancy~\cite{mbnubarcoh}.

In addition topion production results, MiniBooNE and K2K have both published measurements of the axial mass $M_A$\index{axial mass}\index{$M_A$}.  This free parameter of the charged current quasi-elastic (CCQE) cross section appears in the (usually taken as dipole) axial-vector form factor
\begin{equation}
F_{A}(Q^2)=\frac{F_A(0)}{(1+\frac{Q^2}{M_A^2})^2}\ ,
\end{equation}
where $-Q^2=q^2$ is the square of the four-momentum transfer.  With the two-body final state of CCQE scattering, $Q^2$ can be determined from the outgoing lepton's kinematics and the incoming neutrino's direction.  $M_A$ can subsequently be extracted from the observed $Q^2$ distribution.  Two difficulties arise.  The first is that the absolute rate of neutrinos is not well understood.  This is handled by doing a shape-only fit to the $Q^2$ distribution ({\em i.e.}, by assigning an infinite normalization uncertainty).  The second is that nuclear effects are poorly modeled.  K2K excludes the lowest $Q^2$ values (where nuclear effects appear) and MiniBooNE includes empirical nuclear model parameters in the fit to absorb model deficiencies.

\begin{figure}[htb]
\begin{center}
\parbox{0.45\textwidth}{
\begin{tabular}{cc}
Experiment & $M_A$~(GeV)\\\hline
K2K SciFi  & $1.20\pm0.12$ \cite{ma_scifi}\\
K2K SciBar & $1.14\pm0.11$ \cite{ma_scibar}\\
MiniBooNE  & $1.23\pm0.12$ \cite{ma_mb}\\
bubble chambers & $1.03\pm0.02$ \cite{ma_bubble}
\end{tabular}
}
\parbox{0.5\textwidth}{
\epsfig{file=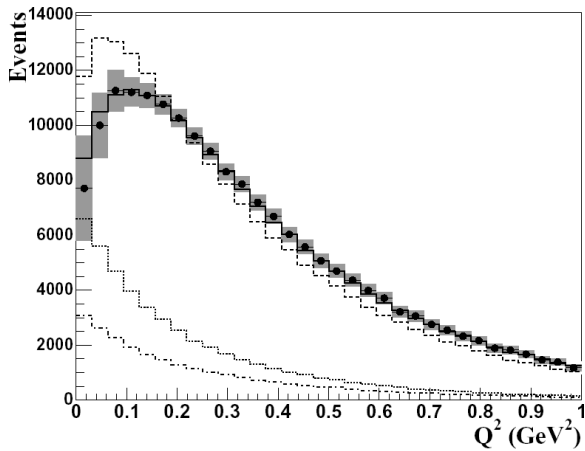,width=0.5\textwidth}
}
\caption{(Left) $M_A$ values from recent measurements by K2K and MiniBooNE along with the average of two decades' worth of bubble chamber experiments.  (Right) The MiniBooNE $Q^2$ distribution.  The points are the data and the shaded bars reflect the systematic errors.  The topmost dashed and solid histograms show the Monte Carlo expectation before and after the $M_A$ and nuclear model fit.  Two classes of non-CCQE background are shown by the lower two histograms.  Figure from Ref.~\cite{ma_mb}.}
\label{fig:mb_qe}
\end{center}
\end{figure}

Figure~\ref{fig:mb_qe} shows the MiniBooNE $Q^2$ fit alongside a table of $M_A$ results.  Note that the bubble chamber experiments yield a significantly lower $M_A$ than the three recent measurements, all on nuclear targets.  So far, this discrepancy evades explanation.

\section{Upcoming}
Two new experiments, SciBooNE and MINER$\nu$A, will help clarify the above inelastic and quasi-elastic situations while reducing the errors on several relevant neutrino cross sections.

SciBooNE~\cite{sciboone}\index{SciBooNE} is a transplant of K2K's SciBar detector into the path of the Fermilab Booster neutrino beam, partway between the proton target and the MiniBooNE detector.  In addition to working in concert with MiniBooNE to improve oscillation sensitivity by acting as a near detector, SciBooNE has over 100k $\nu$ and $\overline{\nu}$ interactions from which to extract cross sections.

The SciBooNE detector consists of a 10-kton, 14\,000-channel scintillator bar tracker followed by an electromagnetic calorimeter and muon range stack.  With its fine-grained tracking, SciBooNE can directly measure the kinematics of recoil nucleons and can distinguish $\nu$ from $\overline{\nu}$ charged current scattering via the displacement of the latter's recoil track.  (The recoil neutron is invisible until it undergoes a hard scatter).  The collaboration is working to produce charged and neutral current pion production and (quasi-)elastic scattering cross sections, with first results expected in the coming year.

MINER$\nu$A~\cite{minerva}\index{MINER$\nu$A}, also at Fermilab, will view the energy-tunable broadband NuMI neutrino beam, typically operating in the 1 to 15~GeV range.  Its highly segmented detector (31\,000 channels) will provide tracking resolutions of 2.5~mm, allowing for excellent final state identification.  An important feature of the detector is the presence of multiple nuclear targets (Fe, Pb, C, and He) allowing nuclear effects and $A$-dependences to be studied.  Data taking is expected to begin in 2009.

Looking further into the future, the NuSOnG\index{NuSOnG} collaboration proposes to convert the to-be-decommissioned Tevatron at Fermilab into a 100~GeV neutrino beam to look for TeV-scale modifications to standard model processes and to provide some degree of LHC complementarity~\cite{nusong}.

Finally, several groups~\cite{snsstancu}~\cite{snsnu}~\cite{clear} look to place neutrino detectors near the Spallation Neutron Source\index{SNS}\index{Spallation Neutron Source} being constructed at Oak Ridge National Laboratory, USA.  A by-product of the neutron production is a $10^{15}\;\nu/s$ pulsed neutrino source with a well-understood decay-at-rest spectrum.  Proposals include low energy cross section measurements, improved sensitivity to LSND-like oscillations, and searches for beyond-the-standard-model components of the weak interaction (via modifications to the $\nu_e$-from-$\mu$-decay energy spectrum).


\begin{thebibliography}{99}


\bibitem{LSND}
  A.~Aguilar {\it et al.}  [LSND Collaboration],
  Phys.\ Rev.\  D {\bf 64}, 112007 (2001)

\bibitem{mbosc}
  A.~A.~Aguilar-Arevalo {\it et al.}  [MiniBooNE Collaboration],
  Phys.\ Rev.\ Lett.\  {\bf 98}, 231801 (2007)

\bibitem{mbcombined}
  A.~A.~Aguilar-Arevalo {\it et al.}  [MiniBooNE Collaboration],
  Phys.\ Rev.\  D {\bf 78}, 012007 (2008)

\bibitem{hhh}
  J.~A.~Harvey, C.~T.~Hill and R.~J.~Hill,
  Phys.\ Rev.\ Lett.\  {\bf 99}, 261601 (2007)

\bibitem{chorus}
  E.~Eskut {\it et al.}  [CHORUS Collaboration],
  Nucl.\ Phys.\  B {\bf 793}, 326 (2008)

\bibitem{lipari}
  P.~Lipari, M.~Lusignoli and F.~Sartogo,
  Phys.\ Rev.\ Lett.\  {\bf 74}, 4384 (1995)

\bibitem{k2kncpi0}
  C.~Mariani [for the K2K Collaboration],
  AIP Conf.\ Proc.\  {\bf 967}, 174 (2007).

\bibitem{k2kccpip}
  A.~Rodriguez {\it et al.}  [K2K Collaboration],
  Phys.\ Rev.\  D {\bf 78}, 032003 (2008)

\bibitem{anl1}
  G.~M.~Radecky {\it et al.},
  Phys.\ Rev.\  D {\bf 25}, 1161 (1982)
  [Erratum-ibid.\  D {\bf 26}, 3297 (1982)].

\bibitem{anl2}
  S.~J.~Barish {\it et al.},
  Phys.\ Rev.\  D {\bf 16}, 3103 (1977).

\bibitem{reinsehgal}
  D.~Rein and L.~M.~Sehgal,
  Nucl.\ Phys.\  B {\bf 223}, 29 (1983).

\bibitem{mbcoh}
  A.~A.~Aguilar-Arevalo {\it et al.}  [MiniBooNE Collaboration],
  Phys.\ Lett.\  B {\bf 664}, 41 (2008)

\bibitem{mbnubarcoh}
  V.~T.~Nguyen,
  arXiv:0806.2347 [hep-ex].

\bibitem{ma_scifi}
  R.~Gran {\it et al.}  [K2K Collaboration],
  Phys.\ Rev.\  D {\bf 74}, 052002 (2006)

\bibitem{ma_scibar}
  K2K Collaboration, in preparation.

\bibitem{ma_mb}
  A.~A.~Aguilar-Arevalo {\it et al.}  [MiniBooNE Collaboration],
  Phys.\ Rev.\ Lett.\  {\bf 100}, 032301 (2008)

\bibitem{ma_bubble}
  V.~Bernard, L.~Elouadrhiri and U.~G.~Meissner,
  J.\ Phys.\ G {\bf 28}, R1 (2002)

\bibitem{sciboone}
  http://www-sciboone.fnal.gov

\bibitem{minerva}
  http://minerva.fnal.gov

\bibitem{nusong}
  http://www-nusong.fnal.gov

\bibitem{snsstancu}
  I.~Stancu,
  AIP Conf.\ Proc.\  {\bf 981}, 34 (2008).

\bibitem{snsnu}
  http://www.phy.ornl.gov/nusns

\bibitem{clear}
  K. Scholberg, APS April Meeting, abstract M13.0009 (2008)

\end{thebibliography}
\end{document}